\documentclass[11pt]{article}
\usepackage{a4,bm,epsfig}
\usepackage{ifthen,array}
\newcommand{\ams}{\usepackage{amsfonts,amssymb,amsmath}}

\ams
\allowdisplaybreaks[3]
\newlength{\textwidthorig}
\newlength{\oddsidemarginorig}
\newlength{\textheightorig}
\newlength{\topmarginorig}
\def\seitenlaengenabsolut#1 #2 #3 #4 {\setlength{\textwidth}{#1}
                                      \setlength{\oddsidemargin}{#2}
                                      \setlength{\textheight}{#3}
                                      \setlength{\topmargin}{#4}}
\def\seitenlaengenrelzustandard#1 #2 #3 #4 {\setlength{\textwidth}{\textwidthorig+#1}
                                            \setlength{\oddsidemargin}{\oddsidemarginorig+#2}
                                            \setlength{\textheight}{\textheightorig+#3}
                                            \setlength{\topmargin}{\topmarginorig+#4}}
\def\seitenlaengenrelzuvorher#1 #2 #3 #4 {\addtolength{\textwidth}{#1}
                                          \addtolength{\oddsidemargin}{#2}
                                          \addtolength{\textheight}{#3}
                                          \addtolength{\topmargin}{#4}}
\newcommand{\standardseite}{\seitenlaengenrelzuvorher2.2cm -0.8cm 1.8cm -1.5cm }   %
\standardseite

\newlength{\laengespatium}

\newcommand{\nach}{\longrightarrow}      

\newcommand{\auf}{\longmapsto}           
\newcommand{\txtauf}[1]{\auf}            
\newcommand{\impliz}{\Longrightarrow}    
\newcommand{\aequ}{\Longleftrightarrow}  
 
\newcommand{\invimpliz}{\Longleftarrow}  
\newcommand{\iso}{\cong}                 
\newcommand{\ident}{\equiv}              
\newcommand{\teilmenge}{\subseteq}       
\newcommand{\obermenge}{\supseteq}       
\newcommand{\aeqrel}{\sim}               
\newcommand{\nichtin}{\not\in}

\newcommand{\leeremenge}{\varnothing}    
\newcommand{\kreuz}{\times}              

\newcommand{\betraganpass}[1]%
           {\left| #1 \right|}           
\newcommand{\bigbetrag}[1]%
           {\bigl|{#1}\bigr|}            
\newcommand{\betrag}[1]%
           {|{#1}|}                      
\newcommand{\betragnichtanpass}[1]%
           {\mid #1 \mid}                
\newcommand{\norm}[1]%
           {{}{\parallel}#1{\parallel}{}}      
\newcommand{\erww}[1]%
           {\langle #1 \rangle}          
\newcommand{\skalprod}[2]%
           {\langle #1,#2 \rangle}       
\newcommand{\supnorm}[1]{{\norm{#1}_\infty}}        
\newcommand{\quer}{\overline}            
\newcommand{\dach}{\widehat}             
\newcommand{\inv}[1]{\frac{1}{#1}}       
\newcommand{\einhalb}{\inv{2}}           
\newcommand{\im}{\text{im\;}}                          
\newcommand{\ido}{\text{id}}                           
\newcommand{\inter}{\text{int}\:}                      
\newcommand{\elanz}{\#}                                
\newcommand{\Hom}{\text{Hom}}                          
\newcommand{\Maps}{\text{Maps}}                        
\newcommand{\field}[1]{\mathbb{#1}}                    
\newcommand{\R}{{\field{R}}}                           
\newcommand{\rnkl}[2]{\raisebox{-0.4ex}{$#1$}%
\raisebox{-0.12ex}{{\large$\setminus$}}\,#2}   
\newcommand{\agb}{{\overline{{\cal A}/{\cal G}}}}      
\newcommand{\agbfact}[1][]{\text{$\agb/\!\aeqrel$}}    
\newcommand{\Ab}{{\overline{{\cal A}}}}                
\newcommand{\Gb}{{\overline{{\cal G}}}}                

\newcommand{\gen}{{\text{gen}}}
\newcommand{\qa}{{\quer{A}}}                           
\newcommand{\qg}{{\quer{g}}}                           

\newcommand{\holgr}{{\mathbf H}}                       
\newcommand{\bz}{{\mathbf B}}                          

\newcommand{\gross}[1]{{\boldsymbol #1}}               
\newcommand{\gc}{\gross{\gamma}}                       
\newcommand{\gd}{\gross{\delta}}                       
\newcommand{\Pf}{{\cal P}}                             
\newcommand{\BB}{\uparrow\uparrow}                     
\newcommand{\hyph}{\upsilon}                           
\newcommand{\Hyph}{Y}                                  
\newcommand{\Haar}{{\text{Haar}}}                      
\newcommand{\LG}{{\mathbf{G}}}                         
\newcommand{\aeqrelzush}[1][]{\sim}                    

\newcommand{\nklza}[1][]{\ifthenelse{\equal{#1}{}}     
                                    {\rnkl{Z(\holgr_\qa)}{\LG}}        
                                   {\rnkl{Z(\holgr_{#1})}{\LG}}}       
\newcommand{\nkla}[1][]{\ifthenelse{\equal{#1}{}}      
                                    {\rnkl{\bz(\qa)}{\Gb}}        
                                    {\rnkl{\bz(#1)}{\Gb}}}       

\newcommand{\YM}{{\text{YM}}}                          

\newcommand{\ymwirk}[1][]{\ifthenelse{\equal{#1}{}}{S_{\YM}}{S_{\YM,#1}}}

\newcommand{\bmat}{\begin{pmatrix}}
\newcommand{\emat}{\end{pmatrix}}

\newcommand{\ListNullAbstaende}{\setlength{\topsep}{0pt}%
                                \setlength{\parskip}{0pt}%
                                \setlength{\partopsep}{0pt}%
                                \setlength{\itemsep}{0pt}%
                                \setlength{\parsep}{0pt}}
\newcommand{\ListNurAnstrichAbstand}{\setlength{\topsep}{0pt}%
                                     \setlength{\parskip}{0pt}%
                                     \setlength{\partopsep}{0pt}%
                                     \setlength{\parsep}{0pt}}
\newenvironment{StandardListe}[2]%
               {\begin{list}%
                      {#1}%
                      {\settowidth{\leftmargin}{M#1}%
                       \settowidth{\labelwidth}{#1}%
                       \settowidth{\labelsep}{M}%
                       #2%
                      }%
                }%
               {\end{list}}%
\newenvironment{EinfachListe}[1]%
               {\begin{StandardListe}{#1}{\ListNullAbstaende}}%
               {\end{StandardListe}}%
               {\begin{StandardListe}{#1}{\ListNurAnstrichAbstand}}%
               {\end{StandardListe}}%
\newcommand{\labelsatz}[1]{#1}
\newcounter{listennr}                      %
\newlength{\hilfslaenge}
\newlength{\stdlabellaenge}
\newlength{\maximum}
\newcommand{\stdlabel}{}
\newcommand{\Maximum}{}
\newcommand{\iitem}[1][]{\ifthenelse{\equal{#1}{}}%
                           {\item \setlength{\hilfslaenge}{\stdlabellaenge}}%
                           {\item[\labelsatz{#1}\hfill]%
                            \settowidth{\hilfslaenge}{\labelsatz{#1}}}%
                         \ifthenelse{\lengthtest{\maximum < \hilfslaenge}}%
                           {\setlength{\maximum}{\hilfslaenge}%
                            \ifthenelse{\equal{#1}{}}%
                               {\renewcommand{\Maximum}{\stdlabel}}%
                               {\renewcommand{\Maximum}{#1}}}%
                           {}%
                      }      
\makeatletter
\newenvironment{AutoLabelLaengenListe}[2][]%
               {\begin{list}%
                      {\labelsatz{#1}\hfill}%
                      {\stepcounter{listennr}%
                       \settowidth{\leftmargin}{M\labelsatz{\ref{listnr\arabic{listennr}}}}%
                       \settowidth{\labelwidth}{\labelsatz{\ref{listnr\arabic{listennr}}}}%
                       \settowidth{\labelsep}{M}%
                       \settowidth{\stdlabellaenge}{\labelsatz{#1}}%
                       \renewcommand{\stdlabel}{#1}%
                       #2%
                       \renewcommand{\Maximum}{}%
                      }%
                }%
               {\renewcommand{\@currentlabel}{\Maximum}%
                \label{listnr\arabic{listennr}}%
                \end{list}%
                }%
\makeatother
\newenvironment{StandardEinrueckung}[2]%
               {\begin{list}%
                      {#1}%
                      {\settowidth{\leftmargin}{M#1}%
                       \settowidth{\labelwidth}{#1}%
                       \settowidth{\labelsep}{M}%
                       #2%
                      }%
                \item}%
               {\end{list}}%
\newenvironment{Einrueckungpur}[1]%
               {\begin{StandardEinrueckung}{#1}{\ListNullAbstaende}}%
               {\end{StandardEinrueckung}}%
\newenvironment{Einrueckung}[1]%
               {\begin{StandardEinrueckung}{#1}{\setlength{\parsep}{0pt}}}%
               {\end{StandardEinrueckung}}%

\makeatletter
\newcommand{\EineNumZeileGleichung}[2][0.5ex]
           {
            
            \vspace{#1} 
            \noindent
            \stepcounter{equation}
            \renewcommand{\@currentlabel}{\arabic{equation}}%
            \phantom{(\arabic{equation})}\hspace*{\fill}
            $\displaystyle{#2}$
            \hspace*{\fill}
            (\arabic{equation})

            \vspace{#1} 
            
           }
\makeatother
\makeatletter
\newcommand{\EineErwNumZeileGleichung}[2][0.5ex]
           {
            
            \vspace{#1} 
            \noindent
            \stepcounter{equation}
            \renewcommand{\@currentlabel}{\arabic{equation}}%
            \phantom{(\arabic{equation})}\hspace*{\fill}
            #2 %
            \hspace*{\fill}
            (\arabic{equation})

            \vspace{#1} 
            
           }
\makeatother
\newcommand{\breitrel}[1]{\hspace*{\tabcolsep} #1 \hspace*{\tabcolsep}}
\newlength{\abstaug}              %
\newenvironment{AllgUnnumGleichung}[2][1.0ex]
               {
  
                \setlength{\abstaug}{#1}
                \vspace{\abstaug}
                \hspace*{\fill}
                $\begin{array}[t]{#2}
                }%
               {\end{array}$
                \hspace*{\fill}
  
                \vspace{\abstaug}

                }%
\newenvironment{AllgNumGleichung}[2][0.0ex]
               {
  
                \setlength{\abstaug}{#1}
                \vspace{\abstaug}
                $\begin{tabular*}{\textwidth}[t]{#2}
                }%
               {\end{tabular*}$

                \vspace{\abstaug}

               }%
\newenvironment{StandardUnnumGleichungKlein}[1][0ex]
               {\renewcommand{\s}{\\[#1] }%
                \begin{AllgUnnumGleichung}{rcl}}%
               {\end{AllgUnnumGleichung}}%
\newenvironment{StandardUnnumGleichung}[1][0ex]%
               {\renewcommand{\s}{\\[#1] }%
                \begin{AllgUnnumGleichung}{>{\displaystyle}rc>{\displaystyle}l}}%
               {\end{AllgUnnumGleichung}}%
\newenvironment{XrelYZNumGleichung}[1][0ex]
               {\renewcommand{\s}{\\[#1] }%
                \begin{AllgNumGleichung}{rcll}}%
               {\end{AllgNumGleichung}}%
\newcommand{\erl}[1]{\hfill\mbox{\hspace*{1.5em}\small (#1)}}

\newcommand{\erllang}[2][0.5\textwidth]%
              {\hfill\hspace*{1.5em}%
               \begin{minipage}[t]{#1}{\small%
                          \begin{list}{(}{\ListNullAbstaende%
                                          \settowidth{\leftmargin}{(}%
                                          \settowidth{\labelwidth}{(}%
                                          \settowidth{\labelsep}{}%
                                         }%
                          \item#2)%
                          \end{list}}%
               \end{minipage}\\[-0.9ex]
              }%
\newcommand{\DefBemUmgeb}[1]%
           {\newenvironment{#1}[1][]%
                           {\begin{Einrueckung}{{\bf #1}}%
                            \ifx##1\empty\else{{\bf ##1}
                            
                                                        }\fi%
                            }%
                           {\end{Einrueckung}}}
\newcommand{\DefSBemUmgeb}[2]
           {\newenvironment{#1}[1][]%
                           {\begin{Einrueckung}{{\bf #2}}%
                            \ifx##1\empty\else{{\bf ##1}
                            
                                                        }\fi%
                            }%
                           {\end{Einrueckung}}}
\makeatletter
\newcommand{\DefBspUmgeb}[3]
           {\newcounter{#2}[#3]%
            \newenvironment{#1}[1][]%
                           {\stepcounter{#2}%
                            \renewcommand{\ZaehlerMarke}{\arabic{#2}}%
                            \renewcommand{\Einzugsname}{{\bf #1 \ZaehlerMarke}}%
                            \begin{Einrueckung}{\Einzugsname}
                            \ifx##1\empty\else{{\bf ##1}\\}\fi%
                            \renewcommand{\@currentlabel}{\ZaehlerMarke}%
                            }%
                           {\end{Einrueckung}}}
\makeatother
\newcommand{\ZaehlerbisEbene}{section}
\newcommand{\Ebenea}{section}
\newcommand{\Ebeneb}{subsection}

\newcommand{\Abschnittnummer}{%
            \ifx\ZaehlerbisEbene\Ebenea{\arabic{section}}%
             \else{%
              \ifx\ZaehlerbisEbene\Ebeneb{\arabic{section}.\arabic{subsection}}%
               \else{\arabic{section}.\arabic{subsection}.\arabic{subsubsection}}%
              \fi}%
            \fi}     
\newcommand{\Abschnittnummerpunkt}{\Abschnittnummer.}     
\newcommand{\Einzugsname}{}
\newcommand{\ZaehlerMarke}{}
\makeatletter
\newcommand{\DefThmUmgeb}[3]%
           {\newcounter{#1}[#3]%
            \newenvironment{#1}[1][]%
                           {\stepcounter{#2}%
                            \setcounter{#1}{\value{#2}}%
                            \renewcommand{\ZaehlerMarke}{\Abschnittnummerpunkt\arabic{#1}}%
                            \renewcommand{\Einzugsname}{{\bf #1 \ZaehlerMarke}}%
                            \begin{Einrueckung}{\Einzugsname}
                            \ifx##1\empty\else{{\bf ##1}
                            
                                                        }\fi%
                            \renewcommand{\@currentlabel}{\ZaehlerMarke}%
                            }%
                           {\end{Einrueckung}}}
\makeatother
\makeatletter
\newcommand{\DefSThmUmgeb}[4]%
           {\newcounter{#1}[#3]%
            \newenvironment{#1}[1][]%
                           {\stepcounter{#2}%
                            \setcounter{#1}{\value{#2}}%
                            \renewcommand{\ZaehlerMarke}{\Abschnittnummerpunkt\arabic{#1}}%
                            \renewcommand{\Einzugsname}{{\bf #4 \ZaehlerMarke}}
                            \begin{Einrueckung}{\Einzugsname}
                            \ifx##1\empty\else{{\bf ##1}

                                                        }\fi%
                            \renewcommand{\@currentlabel}{\ZaehlerMarke}%
                            }%
                           {\end{Einrueckung}}}
\makeatother
\makeatletter
\newcommand{\DefUnterNumThmUmgeb}[5]%
           {\newcounter{#1}[#3]%
            \newcounter{#4}%
            \newenvironment{#1}[1][]%
                           {\ifx##1\empty\else{\stepcounter{#2}\setcounter{#4}{0}}\fi%
                            \stepcounter{#4}%
                            \setcounter{#1}{\value{#2}}%
                            \renewcommand{\ZaehlerMarke}{\Abschnittnummerpunkt\arabic{#1}\alph{#4}}%
                            \renewcommand{\Einzugsname}{{\bf #5 \ZaehlerMarke}}
                            \begin{Einrueckung}{\Einzugsname}
                            \renewcommand{\@currentlabel}{\ZaehlerMarke}%
                            }%
                           {\end{Einrueckung}}}
\makeatother
\newenvironment{Beweis}[1][]%
               {\begin{Einrueckung}{{\bf Beweis}}%
                \ifx#1\empty\else{{\bf #1}

                                            }\fi%
                }%
               {\end{Einrueckung}%
                }%
\newenvironment{Proof}[1][]%
               {\begin{Einrueckung}{{\bf Proof}}%
                \ifx#1\empty\else{{\bf #1}

                                            }\fi%
                }%
               {\end{Einrueckung}%
                }%
               {\begin{Einrueckung}{{\bf \glqq Beweis\grqq}}%
                \ifx#1\empty\else{{\bf #1}
                
                                            }\fi%
                }%
               {\end{Einrueckung}%
                }%
               {\begin{Einrueckung}{{\bf Begr"undung}}%
                \ifx#1\empty\else{{\bf #1}
                
                                            }\fi%
                }%
               {\end{Einrueckung}%
                }%
\newenvironment{Hinrichtung}%
               {\begin{Einrueckungpur}{$\impliz$}}%
               {\end{Einrueckungpur}}%
\newenvironment{Rueckrichtung}%
               {\begin{Einrueckungpur}{$\invimpliz$}}%
               {\end{Einrueckungpur}}%
               {\begin{Einrueckungpur}{\glqq$\teilmenge$\grqq}}%
               {\end{Einrueckungpur}}%
               {\begin{Einrueckungpur}{\glqq$\obermenge$\grqq}}%
               {\end{Einrueckungpur}}%
               {\begin{Einrueckungpur}{"$\teilmenge$"}}%
               {\end{Einrueckungpur}}%
               {\begin{Einrueckungpur}{"$\obermenge$"}}%
               {\end{Einrueckungpur}}%
\newcommand{\qed}{\nopagebreak\hspace*{2em}\hspace*{\fill}{\bf qed}}
\newcommand{\ARabic}{\arabic}
\newcommand{\Nummerntypa}{\arabic}   
\newcommand{\Nummerntypb}{\alph}
\newcommand{\Nummerntypc}{\roman}
\newcommand{\Nummerntypd}{\Alph}

\newcommand{\Nra}{\Nummerntypa{Nummera}}            %
\newcommand{\Nrb}{\Nummerntypb{Nummerb}}            %
\newcommand{\Nrc}{\Nummerntypc{Nummerc}}                
\newcommand{\Nrd}{\Nummerntypd{Nummerd}}                
\newcommand{\ZeichenzuNrTyp}[1]%
           {\ifx#1\ARabic {.}\else{)}%
                  \fi}                              %
\newcommand{\NrZeicha}{\ZeichenzuNrTyp{\Nummerntypa}}
\newcommand{\NrZeichb}{\ZeichenzuNrTyp{\Nummerntypb}}
\newcommand{\NrZeichc}{\ZeichenzuNrTyp{\Nummerntypc}}
\newcommand{\NrZeichd}{\ZeichenzuNrTyp{\Nummerntypd}}
\newcommand{\ListMarkea}%
           {\Nra\NrZeicha}
\newcommand{\ListMarkeb}%
           {\Nra\NrZeicha\Nrb\NrZeichb}
\newcommand{\ListMarkec}%
           {\Nra\NrZeicha\Nrb\NrZeichb\Nrc\NrZeichc}
\newcommand{\ListMarked}%
           {\Nra\NrZeicha\Nrb\NrZeichb\Nrc\NrZeichc\Nrd\NrZeichd}
\newcommand{\Anfangszeichen}{}
\newcommand{\Anfangspunkt}{}
\newcounter{Schachtelebene}
\newcounter{Hilfszaehler}
\newcommand{\Hilfsbefehl}{}
\newcommand{\Schachtelebene}{\alph{Schachtelebene}}
\makeatletter
\newenvironment{AllgNumerierteListe}[2][]
               {\addtocounter{Schachtelebene}{1}%
		\setcounter{Hilfszaehler}{#2}%
                \renewcommand{\Anfangszeichen}%
                             {\renewcommand{\Hilfsbefehl}{\csname Nummerntyp\Schachtelebene \endcsname}%
                              \Hilfsbefehl{Hilfszaehler}}%
                \renewcommand{\Anfangspunkt}%
                             {\csname NrZeich\Schachtelebene \endcsname}%
                \begin{list}%
                      {\stepcounter{Nummer\Schachtelebene}%
                       \csname Nr\Schachtelebene \endcsname
                       \csname NrZeich\Schachtelebene \endcsname
                       }%
                      {\settowidth{\leftmargin}{M\Anfangszeichen\Anfangspunkt}%
                       \settowidth{\labelwidth}{\Anfangszeichen\Anfangspunkt}%
                       \settowidth{\labelsep}{M}%
                       \setlength{\topsep}{0pt}%
                       \setlength{\parskip}{0pt}%
                       \setlength{\partopsep}{0pt}%
                       \setlength{\itemsep}{0pt}%
                       \setlength{\parsep}{0pt}%
                      }%
                \renewcommand{\@currentlabel}{\csname ListMarke\Schachtelebene \endcsname}%
                }%
               {\ifthenelse{\equal{}{}}{\setcounter{Nummer\Schachtelebene}{0}}{}
                \addtocounter{Schachtelebene}{-1}%
                \end{list}}
\makeatother
\newenvironment{NumerierteListe}[1]
               {\begin{AllgNumerierteListe}{#1}}
               {\end{AllgNumerierteListe}}
\newenvironment{WeiterNumerierteListe}[1]
               {\begin{AllgNumerierteListe}[Weiter]{#1}}
               {\end{AllgNumerierteListe}}

\newcommand{\UnnumAnfangszeichen}{}
\newcounter{UnnumSchachtelebene}
\newcommand{\UnnumSchachtelebene}{\alph{UnnumSchachtelebene}}
\makeatletter
\newenvironment{UnnumerierteListe}%
               {\addtocounter{UnnumSchachtelebene}{1}%
                \renewcommand{\UnnumAnfangszeichen}%
                             {\csname UnnumZeich\UnnumSchachtelebene \endcsname}%
                \begin{list}%
                      {\UnnumAnfangszeichen}%
                      {\settowidth{\leftmargin}{M\UnnumAnfangszeichen}%
                       \settowidth{\labelwidth}{\UnnumAnfangszeichen}%
                       \settowidth{\labelsep}{M}%
                       \setlength{\topsep}{0pt}%
                       \setlength{\parskip}{0pt}%
                       \setlength{\partopsep}{0pt}%
                       \setlength{\itemsep}{0pt}%
                       \setlength{\parsep}{0pt}%
                      }%
                }%
               {\addtocounter{UnnumSchachtelebene}{-1}%
                \end{list}}
\makeatother
\newlength{\fktdefhilfslaenge}
\newcommand{\ohnefktdef}[4]
           {\hspace*{\fill}
            $\begin{array}[t]{ccc}%
            #1 & \nach & #2 \\
            #3 & \auf  & #4
            \end{array}$
            \hspace*{\fill}}
\newcommand{\fktdef}[5]
           {\hspace*{\fill}
            $\begin{array}[t]{cccc}%
            #1: & #2 & \nach & #3 \\    
                & #4 & \auf  & #5
            \end{array}$
            \settowidth{\fktdefhilfslaenge}{$#1$:}
            \hspace*{0.6 \fktdefhilfslaenge}  
            \hspace*{\fill}}
\newcommand{\fktdefpur}[5]
           {$\begin{array}[t]{cccc}%
            #1: & #2 & \nach & #3 \\    
                & #4 & \auf  & #5
            \end{array}$}
\newcommand{\fktdefabgesetztpur}[5]
           {
            
            $\begin{array}[t]{cccc}%
            #1: & #2 & \nach & #3 \\    
                & #4 & \auf  & #5
            \end{array}$
            \settowidth{\fktdefhilfslaenge}{$#1$:}
            \hspace*{0.6 \fktdefhilfslaenge}
            
           }
\newcommand{\fktdefabgesetzt}[5]
           {
           
            \hspace*{\fill}
            $\begin{array}[t]{cccc}%
            #1: & #2 & \nach & #3 \\    
                & #4 & \auf  & #5
            \end{array}$
            \settowidth{\fktdefhilfslaenge}{$#1$:}
            \hspace*{0.6 \fktdefhilfslaenge}  
            \hspace*{\fill}
            
            }
\newcommand{\ohnefktdefabgesetzt}[4]
           {      

            \hspace*{\fill}
            $\begin{array}[t]{ccc}%
            #1 & \nach & #2 \\
            #3 & \auf  & #4
            \end{array}$
            \hspace*{\fill}

            }
\newcommand{\doppelohnefktdefabgesetzt}[6]
           {

            \hspace*{\fill}
            $\begin{array}[t]{ccccc}%
            #1 & \nach & #2 & \nach & #3\\
            #4 & \auf  & #5 & \auf  & #6
            \end{array}$
            \hspace*{\fill}

            }
\newcommand{\anhang}%
           {\appendix
            \sectioninh{Anhang}
            \renewcommand{\Abschnittnummer}{%
                  \ifx\ZaehlerbisEbene\Ebenea{\Alph{section}}%
                  \else{%
                        \ifx\ZaehlerbisEbene\Ebeneb{\Alph{section}.\arabic{subsection}}%
                        \else{\Alph{section}.\arabic{subsection}.\arabic{subsubsection}}%
                        \fi}%
                  \fi}%
            \renewcommand{\Abschnittnummerpunkt}{\Abschnittnummer.}     
            }            
\newcommand{\anhangengl}%
           {\appendix
            \sectioninh{Appendix}
            \renewcommand{\Abschnittnummer}{%
                  \ifx\ZaehlerbisEbene\Ebenea{\Alph{section}}%
                  \else{%
                        \ifx\ZaehlerbisEbene\Ebeneb{\Alph{section}.\arabic{subsection}}%
                        \else{\Alph{section}.\arabic{subsection}.\arabic{subsubsection}}%
                        \fi}%
                  \fi}%
            \renewcommand{\Abschnittnummerpunkt}{\Abschnittnummer.}     
            }

\newcounter{wdhlstufe}
\newcommand{\sectioninh}[1]%
           {\section*{#1}%
            \addcontentsline{toc}{section}{#1}}
\newcommand{\bezeichnung}[3]%
           {\begin{Einrueckungpur}{\hbox to 6em{#1}\hbox to 2.4em{\hfill#2}}
            #3
            \end{Einrueckungpur}}

\newcommand{\doppelteinfach}{e}

\newcommand{\ifdoppelt}[1]{\ifthenelse{\equal{\doppelteinfach}{d}}{#1}{}}
\newcommand{\ifeinfach}[1]{\ifthenelse{\equal{\doppelteinfach}{e}}{#1}{}}

\newlength{\querfhilfsl}              %

\newlength{\hll}

\DefThmUmgeb{Theorem}{Theorem}{\ZaehlerbisEbene}
\DefThmUmgeb{Definition}{Definition}{\ZaehlerbisEbene}
\DefThmUmgeb{Satz}{Theorem}{\ZaehlerbisEbene}
\DefThmUmgeb{Proposition}{Theorem}{\ZaehlerbisEbene}
\DefThmUmgeb{Lemma}{Theorem}{\ZaehlerbisEbene}
\DefThmUmgeb{Folgerung}{Theorem}{\ZaehlerbisEbene}
\DefThmUmgeb{Corollary}{Theorem}{\ZaehlerbisEbene}
\DefThmUmgeb{Vorschrift}{Definition}{\ZaehlerbisEbene}
\DefThmUmgeb{Construction}{Definition}{\ZaehlerbisEbene}
\DefSThmUmgeb{FormSatz}{Theorem}{\ZaehlerbisEbene}{\glqq Satz\grqq} 
\DefThmUmgeb{Vermutung}{Theorem}{\ZaehlerbisEbene}
\DefThmUmgeb{Conjecture}{Theorem}{\ZaehlerbisEbene}
\DefThmUmgeb{Konvention}{Definition}{\ZaehlerbisEbene}
\DefThmUmgeb{Feststellung}{Theorem}{\ZaehlerbisEbene}
\DefUnterNumThmUmgeb{DefinitionZusatzNum}{Definition}{\ZaehlerbisEbene}{DefZN}{Definition}
\DefBspUmgeb{Beispiel}{Beispiel}{subsubsection}
\DefBspUmgeb{Example}{Example}{subsubsection}
\DefBspUmgeb{Frage}{Frage}{section}
\DefBspUmgeb{Question}{Question}{section}
\DefBspUmgeb{Aufgabe}{Aufgabe}{section}
\DefBspUmgeb{Ziel}{Ziel}{section}
\DefBemUmgeb{Bemerkung}
\DefBemUmgeb{Remark}
\DefSBemUmgeb{OffeneFrage}{Offene Frage}
\newcommand{\bdf}{\begin{Definition}}
\newcommand{\edf}{\end{Definition}}
\newcommand{\bvorsch}{\begin{Vorschrift}}
\newcommand{\evorsch}{\end{Vorschrift}}
\newcommand{\bconst}{\begin{Construction}}
\newcommand{\econst}{\end{Construction}}
\newcommand{\bthm}{\begin{Theorem}}
\newcommand{\ethm}{\end{Theorem}}
\newcommand{\bsatz}{\begin{Satz}}
\newcommand{\esatz}{\end{Satz}}
\newcommand{\bprop}{\begin{Proposition}}
\newcommand{\eprop}{\end{Proposition}}
\newcommand{\blem}{\begin{Lemma}}
\newcommand{\elem}{\end{Lemma}}
\newcommand{\bfolg}{\begin{Folgerung}}
\newcommand{\efolg}{\end{Folgerung}}
\newcommand{\bcorr}{\begin{Corollary}}
\newcommand{\ecorr}{\end{Corollary}}
\newcommand{\bfest}{\begin{Feststellung}}
\newcommand{\efest}{\end{Feststellung}}
\newcommand{\bbew}{\begin{Beweis}}
\newcommand{\ebew}{\end{Beweis}}
\newcommand{\bpf}{\begin{Proof}}
\newcommand{\epf}{\end{Proof}}
\newcommand{\bwnum}{\begin{WeiterNumerierteListe}}
\newcommand{\ewnum}{\end{WeiterNumerierteListe}}
\newcommand{\bdfzn}{\begin{DefinitionZusatzNum}}
\newcommand{\edfzn}{\end{DefinitionZusatzNum}}
\newcommand{\bbem}{\begin{Bemerkung}}
\newcommand{\ebem}{\end{Bemerkung}}
\newcommand{\brem}{\begin{Remark}}
\newcommand{\erem}{\end{Remark}}
\newcommand{\bnum}{\begin{NumerierteListe}}
\newcommand{\enum}{\end{NumerierteListe}}
\newcommand{\bunum}{\begin{UnnumerierteListe}}
\newcommand{\eunum}{\end{UnnumerierteListe}}
\newcommand{\bbsp}{\begin{Beispiel}}
\newcommand{\ebsp}{\end{Beispiel}}
\newcommand{\bex}{\begin{Example}}
\newcommand{\eex}{\end{Example}}
\newcommand{\bfrag}{\begin{Frage}}
\newcommand{\efrag}{\end{Frage}}
\newcommand{\bquest}{\begin{Question}}
\newcommand{\equest}{\end{Question}}
\newcommand{\baufg}{\begin{Aufgabe}}
\newcommand{\eaufg}{\end{Aufgabe}}
\newcommand{\bof}{\begin{OffeneFrage}}
\newcommand{\eof}{\end{OffeneFrage}}
\newcommand{\bverm}{\begin{Vermutung}}
\newcommand{\everm}{\end{Vermutung}}
\newcommand{\bconj}{\begin{Conjecture}}
\newcommand{\econj}{\end{Conjecture}}
\newcommand{\bkonv}{\begin{Konvention}}
\newcommand{\ekonv}{\end{Konvention}}
\newcommand{\bglklein}{\begin{StandardUnnumGleichungKlein}}
\newcommand{\eglklein}{\end{StandardUnnumGleichungKlein}}
\newcommand{\bgl}{\begin{StandardUnnumGleichung}}
\newcommand{\egl}{\end{StandardUnnumGleichung}}
\newcommand{\bglrtext}{\begin{XrelYZNumGleichung}}
\newcommand{\eglrtext}{\end{XrelYZNumGleichung}}

\newcommand{\berlgl}{\begin{StandardUnnumGleichung}}
\newcommand{\eerlgl}{\end{StandardUnnumGleichung}}
\newcommand{\beinrueck}{\begin{Einrueckungpur}} 
\newcommand{\eeinrueck}{\end{Einrueckungpur}}
\newcommand{\beinflist}{\begin{EinfachListe}} 
\newcommand{\eeinflist}{\end{EinfachListe}}
\newcommand{\beq}{\begin{equation}}
\newcommand{\eeq}{\end{equation}}
\newcommand{\bhin}{\begin{Hinrichtung}}
\newcommand{\ehin}{\end{Hinrichtung}}
\newcommand{\brueck}{\begin{Rueckrichtung}}
\newcommand{\erueck}{\end{Rueckrichtung}}
\newcommand{\bvl}{\begin{AutoLabelLaengenListe}{\ListNullAbstaende}}
\newcommand{\evl}{\end{AutoLabelLaengenListe}}
\newcommand{\df}[1]{{\bf #1}}

\newlength{\adressabstand}
\newcommand{\Bigbetrag}[1]%
           {\Bigl|{#1}\Bigr|}

\newcommand{\diffeo}{\varphi}%
\newcommand{\qfa}{\Theta}%
\newcommand{\ausl}{-}%
\newcommand{\einl}{+}%
\newcommand{\bound}{{\cal{B}}}%
\newcommand{\Pfgen}{\Pf_\gen}
\newcommand{\Grapho}{{\mathrm{Grapho}}}%
\newcommand{\adm}{{\cal Q}}

\newcommand{\qwert}{\lambda}%
\newcommand{\rrr}{\kappa}%
\newcommand{\Germ}{{\mathrm{Germ}}}%
\newcommand{\qgerm}{\rho}%
\newcommand{\neuzh}[1]{\dach{#1}}
               {\addtocounter{Schachtelebene}{1}%
		\setcounter{Hilfszaehler}{#1}%
                \renewcommand{\Anfangszeichen}%
                             {\renewcommand{\Hilfsbefehl}{\csname Nummerntyp\Schachtelebene \endcsname}%
                              \Hilfsbefehl{Hilfszaehler}}%
                \renewcommand{\Anfangspunkt}%
                             {\csname NrZeich\Schachtelebene \endcsname}%
                \begin{list}%
                      {\stepcounter{Nummer\Schachtelebene}%
                       \csname Nr\Schachtelebene \endcsname
                       \csname NrZeich\Schachtelebene \endcsname
                       }%
                      {\settowidth{\leftmargin}{M\Anfangszeichen\Anfangspunkt}%
                       \settowidth{\labelwidth}{\Anfangszeichen\Anfangspunkt}%
                       \settowidth{\labelsep}{M}%
                       \setlength{\topsep}{0pt}%
                       \setlength{\parskip}{0pt}%
                       \setlength{\partopsep}{0pt}%
                       \setlength{\itemsep}{0pt}%
                       \setlength{\parsep}{0pt}%
                      }%
                \renewcommand{\@currentlabel}{\csname ListMarke\Schachtelebene \endcsname}%
                }%
               {\addtocounter{Schachtelebene}{-1}%
                \end{list}}
\newcommand{\extrazeile}[1][]{\enlargethispage{#1\baselineskip}}
\newcommand{\neueseite}{\newpage}
\begin{document}
\title{Construction of Generalized Connections}
\author{Christian Fleischhack\thanks{e-mail: 
            {\tt chfl@mis.mpg.de}} \\   
        \\
        {\normalsize\em Max-Planck-Institut f\"ur Mathematik in den
                        Naturwissenschaften}\\[\adressabstand]
        {\normalsize\em Inselstra\ss e 22--26}\\[\adressabstand]
        {\normalsize\em 04103 Leipzig, Germany}
        \\[-25\adressabstand]      
        {\normalsize\em Institute for Gravitational Physics and Geometry}\\[\adressabstand]
        {\normalsize\em 320 Osmond Lab}\\[\adressabstand]
        {\normalsize\em Penn State University}\\[\adressabstand]
        {\normalsize\em University Park, PA 16802}
        \\[-25\adressabstand]}      
\date{September 4, 2005}
\maketitle
\begin{abstract}
We present a construction method for mappings
between generalized connections, 
comprising, e.g., the action of gauge transformations,
diffeomorphisms and Weyl transformations.
Moreover, criteria for continuity and measure preservation are stated.
\end{abstract}

\section{Introduction}
Generalized (or distributional) connections arise naturally when
attempting the loop quantization of canonical gravity or other gauge
field theories. Often
mappings between such connections are used. Examples are gauge
transformations, diffeomorphisms and Weyl transformations. 
Distributional connections are given as homomorphisms from
the groupoid of paths in the base manifold of a principal fibre bundle
to its structure group. So one typically
tries to define transformed connections 
by modifying the parallel transports of a given one,
path by path. This is not always directly possible. Usually one has to
break paths down to ``simple'' pieces, where the mapping can be defined
more easily.
Afterwards, one patches the simple parts together by homomorphy. However,
here one has to take care of the well-definedness. At the end, one
is interested in the properties of these mappings, in particular,
continuity and measure preservation.

It turns out that most of the transformations considered until now
in that framework,
follow this pattern. So, in the examples listed above, 
the parallel transport along a ``simple'' path is always given
by the parallel transports along some possibly other ``simple'' path
plus some conjugation with structure group elements specifying the
transformation.
Since, therefore, proofs often are very similar for different 
transformations, we are now going to somewhat unify the treatment
in the present paper. We start with some results on the
decomposition of paths in Section \ref{sect:complete} and then
establish the general construction in Section \ref{sect:construction}.
After applying it to connections as in Section \ref{sect:conn} and
introducing the notion of graphical morphisms in Section \ref{sect:grapho},
the main results are presented in Section \ref{sect:main}.
We close the paper with a bunch of
examples in Section \ref{sect:examples}.


Finally, let us fix some manifold $M$ and some connected Lie group $\LG$.
If we make any statements on measures, we will assume $\LG$ to be compact.

\section{Completeness}
\label{sect:complete}
Recall \cite{paper2+4} that a path is a piecewise $C^r$ map from $[0,1]$
to our fixed manifold $M$. Here, the fixed $r$ 
is either a positive integer, $\infty$ or $\omega$. Moreover, we
decide whether we 
restrict ourselves to piecewise embedded paths or not.
A path is said to be trivial iff its image is a single point.
The inverse path $\gamma^{-1}$ of a path $\gamma$ is given by 
$\gamma^{-1}(t) := \gamma(1-t)$.
Two paths $\gamma_1$ and $\gamma_2$ are composable iff the end point 
$\gamma_1(1)$ of the first one coincides with the starting point $\gamma_2(0)$
of the second one. If they are composable, their product is given 
by 
\bgl
(\gamma_1 \gamma_2)(t) 
       \breitrel{:=} \begin{cases}
           \gamma_1(2t)   & \text{ for $t\in[0,\einhalb]$} \\
           \gamma_2(2t-1) & \text{ for $t\in[\einhalb,1]$}
           \end{cases}.
\egl\noindent
An edge $e$ is a path having no self-intersections, i.e., 
from $e(t_1) = e(t_2)$ follows that $\betrag{t_1-t_2}$ is either $0$ or $1$.
Two paths $\gamma_1$ and $\gamma_2$ 
coincide up to the parametrization iff there is some orientation preserving
piecewise 
$C^r$ diffeomorphism 
$\phi: [0,1] \nach [0,1]$, such that $\gamma_1 = \gamma_2 \circ \phi$.
A path is called finite iff it equals up to the parametrization
a finite product of edges and trivial paths.
In what follows, every path will be assumed to be finite.
Next, two paths are equivalent iff there is a finite sequence of paths,
such that two subsequent paths coincide up to the parametrization or
up to insertion or deletion of retracings $\delta\delta^{-1}$. 
This means, that, e.g.,
$\gamma_1\gamma_2$ is equivalent to $\gamma_1 \delta \delta^{-1} \gamma_2$
for all paths $\gamma_1$, $\gamma_2$ and $\delta$.
Finally, we denote the set of all paths by $\Pfgen$, that of all equivalence 
classes of paths by $\Pf$. $\Pf$ is a groupoid.

\bdf
Let $\gamma$ be some path. 

Then a finite 
sequence $\gc := (\gamma_1,\ldots,\gamma_n)$ in $\Pfgen$ is called
\df{decomposition} of $\gamma$ iff $\gamma_1 \cdots \gamma_n$ equals
$\gamma$ 
up to the parametrization.%
\edf
This definition is well defined, since $\gamma_1 (\gamma_2\gamma_3)$
equals $(\gamma_1\gamma_2) \gamma_3$ up to the parametrization.
Moreover, observe that every reparametrization of $\gamma$ gives
a decomposition of $\gamma$.

If confusion is unlikely, we identify $\gamma_1\cdots\gamma_n$ 
and $(\gamma_1,\ldots,\gamma_n)$.

\bdf
Let $\gc := \gamma_1\cdots\gamma_I$ and 
$\gd := \delta_1\cdots\delta_J$
be decompositions of some path $\gamma$.

Then $\gc$ is a \df{refinement} of
$\gd$ iff there are
$0 = I_0 < I_1 < \ldots < I_J = I$, such that 
$\gamma_{I_{j-1}+1} \cdots \gamma_{I_j}$ is a decomposition of 
$\delta_j$ for all $j = 1, \ldots, J$.
We write $\gc \geq \gd$ iff $\gc$ is a refinement of
$\gd$.
\edf

\blem
Let $\gamma$ be some path.

Then the set of all decompositions of $\gamma$ is directed w.r.t.\ $\geq$.
\elem
\bpf
Let $\gc = \gamma_1 \cdots \gamma_I$ be a decomposition of $\gamma$,
i.e., 
$(\ldots((\gamma_1 \gamma_2) \gamma_3) \cdots )\gamma_I$
equals $\gamma \circ \phi^{-1}$
for some piecewise $C^r$ diffeomorphism $\phi$ from $[0,1]$ onto itself.
Now, the nontrivial end points 
of the $\gamma_i$ correspond to the parameter values
$\phi(\inv{2^{I-1}})$, \ldots, $\phi(\inv4)$, and $\phi(\inv2)$
in $\gamma$. In other words, these parameter
values decompose $\gamma$ into the $\gamma_i$.%
\footnote{Note that these values need not be uniquely determined. In fact,
if there is some interval in $[0,1]$, where $\gamma$ 
is constant, then $\phi$ is not uniquely determined by $\gamma$ and $\gc$,
since every $\widetilde\phi$ coinciding with $\phi$ outside that interval
gives $\gamma \circ \phi^{-1} = \gamma \circ \widetilde \phi^{-1}$.}

Let now $\gc_1$ and $\gc_2$ be two decompositions of $\gamma$. Then we
may find two sets of parameter values that decompose $\gamma$ according
to $\gc_1$ and $\gc_2$, respectively. Now, 
decompose $\gamma$ according to the union of these two sets. This gives 
a decomposition $\gc$ of $\gamma$. 
It is easy to check that $\gc$ is a refinement of
both $\gc_1$ and $\gc_2$.
\qed
\epf

\bdf
A subset $\adm$ of $\Pfgen$ is called \df{hereditary} iff 
for each $\gamma\in\adm$
\bnum2
\item
the inverse of $\gamma$ is in $\adm$ again, and
\item
every decomposition of $\gamma$ consists of paths in $\adm$.
\enum
\edf

\bdf
A subset $\adm$ of $\Pfgen$ is called
\df{complete} iff it is hereditary 
and every path in $\Pfgen$ has a decomposition into paths 
in $\adm$.%

\edf
A decomposition consisting of paths in $\adm$ only, will be called
$\adm$-decomposition.

\blem
The set of all edges and trivial paths in $\Pfgen$ is complete.
\elem
\bpf
Clear from the definition of $\Pfgen$.
\qed
\epf

\section{Construction}
\label{sect:construction}

\bdf
\label{def:germ}
Let $\adm$ be some hereditary subset of $\Pfgen$. 

Then a map $\qgerm : \adm \nach \LG$ is called
\df{$\adm$-germ} iff for all $\gamma\in\adm$
\bnum2
\item
\label{punkt:inv(qgerm)}
$\qgerm(\gamma^{-1}) = \qgerm(\gamma)^{-1}$, and
\item
\label{punkt:homom(qgerm)}
$\qgerm(\gamma) = \qgerm(\gamma_1) \qgerm(\gamma_2)$
for all decompositions $\gamma_1\gamma_2$ of $\gamma$.
\enum
The set of all $\adm$-germs from $\adm$ to $\LG$
is denoted by $\Germ(\adm,\LG)$.
\edf
Observe that $\qgerm(\gamma)$ and $\qgerm(\delta)$ coincide if 
$\gamma$ and $\delta$ coincide up to the parametrization. In fact, since
every decomposition $\gamma_1 \gamma_2$ of $\gamma$ is also some
for $\delta$, we may apply property \ref{punkt:homom(qgerm)} above.

Note that we will shortly speak about germs instead of $\adm$-germs,
provided the domain $\adm$ is clear from the context.

\bprop
\label{prop:konstr(zush)}
Let $\adm$ be some complete subset of $\Pfgen$,
and 
let $\qgerm: \adm \nach \LG$ be a germ.

Then we have:
\bunum
\item
There is a unique germ $\neuzh\qgerm : \Pfgen \nach \LG$
extending $\qgerm$.

\item
The map $\neuzh\qgerm$ is given by
\bglklein
\neuzh\qgerm(\gamma) & = & \prod_{i=1}^I \qgerm(\gamma_i)
\eglklein
for each $\gamma\in\Pfgen$, where 
$\gamma_1\cdots\gamma_n$ is any%
\footnote{Recall that, by completeness of $\adm$, 
such a decomposition exists always.}
$\adm$-decomposition of $\gamma$.
\item
The map $\neuzh\qgerm$ is constant on equivalence classes in $\Pfgen$.
\item
The induced map
$[\neuzh\qgerm] : \Pf \nach \LG$ is a homomorphism. 
\eunum
\eprop

\bpf
Let us first define the desired map $\neuzh\qgerm$ as given in the 
proposition above and now check its properties.
\bnum{3}
\item
$\neuzh\qgerm$ does not depend on the choice of the 
$\adm$-decomposition.

Let $\gc$ and $\gd$ be two $\adm$-decompositions of $\gamma$.
Since, by assumption, 
every path in $\adm$ has $\adm$-decompositions only, and
since the set of decompositions of a path is directed w.r.t.\ $\geq$,
we may assume $\gc \geq \gd$.
But, in this case the well-definedness follows directly 
from the definitions and germ property \ref{punkt:homom(qgerm)} 
of $\qgerm$.
\item
$\neuzh\qgerm$ is constant on equivalence classes in $\Pfgen$.

Let $\gamma$ and $\delta$ in $\Pfgen$ be equivalent. By definition,
it is sufficient to check the following two cases:
\bunum
\item
$\gamma$ and $\delta$ coincide up to the parametrization.

Since every $\adm$-decomposition of $\gamma$ is also one of $\delta$,
we have $\qgerm(\gamma) = \qgerm(\delta)$.
\item
There is some $\varepsilon$ in $\Pfgen$ and some decomposition 
$\gamma_1 \gamma_2$ of $\gamma$, such that 
$\delta$ equals the product 
of $\gamma_1$, $\varepsilon$, $\varepsilon^{-1}$ and $\gamma_2$.

Now, in this case, 
choose some $\adm$-decompositions $\varepsilon_1 \cdots \varepsilon_K$
of $\varepsilon$ and $\gamma_{s1}\cdots\gamma_{sI_s}$ of $\gamma_s$ with
$s=1,2$. Then 
$\gamma_{11}\cdots\gamma_{1I_1} \: \gamma_{21}\cdots\gamma_{2I_2}$
is a $\adm$-decomposition of $\gamma$ and
$\gamma_{11}\cdots\gamma_{1I_1} \: \varepsilon_1 \cdots \varepsilon_K \:
 \varepsilon_K^{-1} \cdots \varepsilon_1^{-1} \: \gamma_{21}\cdots\gamma_{2I_2}$
one of $\delta$.
Hence, we have
\bglklein
\neuzh\qgerm(\delta) 
  & = & \qgerm(\gamma_{11}) \cdots \qgerm(\gamma_{1I_1}) \: 
         \qgerm(\varepsilon_1) \cdots \qgerm(\varepsilon_K) \:
\\&&\hspace*{3em}
         \qgerm(\varepsilon_K^{-1}) \cdots \qgerm(\varepsilon_1^{-1}) \: 
         \qgerm(\gamma_{21}) \cdots \qgerm(\gamma_{2I_2}) 
        \erl{Definition of $\neuzh\qgerm$} \\
  & = & \qgerm(\gamma_{11}) \cdots \qgerm(\gamma_{1I_1}) \: 
         \qgerm(\gamma_{21}) \cdots \qgerm(\gamma_{2I_2}) 
        \erl{Property \ref{punkt:inv(qgerm)} of $\qgerm$} \\
  & = & \neuzh\qgerm(\gamma).
        \erl{Definition of $\neuzh\qgerm$}
\eglklein
\eunum
\item
$\neuzh\qgerm$ is a germ extending $\qgerm$, 
and $[\neuzh\qgerm]$ is a homomorphism.

This is proven as the statements above.
\item
$\neuzh\qgerm$ is the only germ extending $\qgerm$.

If $\neuzh\qgerm'$ is some other germ extending $\qgerm$ different
from $\neuzh\qgerm$, then there is some $\gamma\in\Pfgen$ with 
$\neuzh\qgerm'(\gamma) \neq \neuzh\qgerm(\gamma)$. Now, choose
a $\adm$-decomposition $\gamma_1 \cdots \gamma_I$ of $\gamma$. By
the properties of a germ, there is some $i$ with
$\neuzh\qgerm'(\gamma_i) \neq \neuzh\qgerm(\gamma_i)$. However, since both
$\neuzh\qgerm'$ and $\neuzh\qgerm$ extend $\qgerm$, both sides
are equal to $\qgerm(\gamma_i)$. Contradiction.
\qed
\enum
\epf

\section{Connections}
\label{sect:conn}
To be prepared for the main results of this paper, let us briefly
recall \cite{a28,paper2+4,paper3}
the basic definitions and properties of generalized connections.
Algebraically, the space%
\footnote{The elements of $\Ab$ are denoted by $\qa$ or, synonymously,
by $h_\qa$.} 
$\Ab$ of generalized connections equals 
$\Hom(\Pf,\LG)$. To equip $\Ab$ with a topology and
measures thereon, we have to go again into the field of paths.
Segments of a path are restrictions of that path to connected subintervals,
affinely stretched to maps with domain $[0,1]$.
Initial and final segments of paths are defined naturally.
We will write $\gamma_1\BB\gamma_2$ iff there is some path $\gamma$ being
(possibly up to the parametrization) an initial segment of both $\gamma_1$
and $\gamma_2$.
A hyph $\hyph$ is now some finite collection 
$(\gamma_1,\ldots,\gamma_n)$ of edges each having a ``free'' point. 
This means, for at least one direction none of the segments of $\gamma_i$ 
starting in that point in this direction, is (up to the parametrization)
a full segment
of some of the $\gamma_j$ with $j<i$. 
The decomposition of paths and the inclusion relation generate
a directed ordering on the set of hyphs.
Now,
\bglklein
\Ab \breitrel\ident \Hom(\Pf,\LG) \breitrel= \varprojlim_\hyph \Ab_\hyph,
\eglklein\noindent
with $\Ab_\hyph := \Hom(\Pf_\hyph,\LG) \iso \LG^{\elanz\hyph}$ 
given the topology induced by that of $\LG$.
Here, $\Pf_\hyph$ is the subgroupoid of $\Pf$, generated freely by the 
(equivalence classes of the) edges in $\hyph$. Then 
$\pi_\hyph : \Ab \nach \LG^{\elanz\hyph}$ 
with $\pi_\hyph(\qa) := \qa([\hyph])$ is always continuous.

\bprop
\label{prop:qfa_cont}
Let $\adm$ be some complete subset of $\Pfgen$.
Let $X$ be some topological space, and let $\qwert : X \nach \Germ(\adm,\LG)$
be some map.
Finally, assume that the map $\bigl(\qwert(\cdot)\bigr)(\gamma) : X \nach \LG$
is continuous for all $\gamma\in\adm$.

Then 
\fktdefabgesetzt{\qfa_\qwert}{X}{\Ab}{x}{[\neuzh{\qwert(x)}]}
is continuous, 
where $\:\neuzh\cdot\:$ is given as in Proposition \ref{prop:konstr(zush)}.
\eprop
\bpf
It is sufficient \cite{paper2+4} to prove that 
$\pi_{\gamma} \circ \qfa_\qwert : X \nach \LG$
is continuous for all edges $\gamma$. Since the multiplication in $\LG$
is continuous and $\adm$ is complete, 
we even may restrict ourselves to the cases 
of $\gamma \in \adm$.
Here, however, the assertion
follows immediately from
\bgl
(\pi_{\gamma} \circ \qfa_\qwert) (x) 
   & \ident & \pi_{\gamma} ([\neuzh{\qwert(x)}]) 
   \breitrel= [\neuzh{\qwert(x)}]([\gamma]) 
   \breitrel= \neuzh{\qwert(x)}(\gamma) 
   \breitrel\ident \qwert(x)(\gamma),
\egl
i.e., $\pi_{\gamma} \circ \qfa_\qwert = \bigl(\qwert(\cdot)\bigr)(\gamma)$ 
for all $\gamma\in\adm$.
\qed
\epf

We close with
\blem
\label{lem:coinc_krit}
Two generalized connections coincide iff they coincide for all
(equivalence classes of) paths of a complete subset of $\Pfgen$.
\elem

\section{Graphomorphisms}
\label{sect:grapho}
Among the possibilities to modify connections, i.e.,
mappings from $\Pf$ to $\LG$, those induced by 
transformations of $\Pf$ are very
important. In particular, they arise in the context of diffeomorphisms
that naturally induce an action on paths and graphs.
But not only diffeomorphisms give nicely behaving transformations
of paths and graphs.
Hence, we extend the notion of diffeomorphisms.

\bdf
Let $\diffeo : M \nach M$ be a map.
\bunum
\item
$\diffeo$ is called \df{graphical homomorphism} iff $\diffeo$ 
induces%
\footnote{This includes that $\diffeo \circ \gamma$ is in $\Pfgen$
for all $\gamma\in\Pfgen$.}
a groupoid homomorphism on $\Pf$.
\item
$\diffeo$ is called \df{graphical isomorphism} 
(or shorter:\ \df{graphomorphism})
iff $\diffeo$ is bijective and
both $\diffeo$ and $\diffeo^{-1}$ are graphical homomorphisms.

\eunum
The set of all graphomorphisms is denoted by $\Grapho(M)$.
\edf
Of course, each diffeomorphism is a graphomorphism.

For technical purposes, it is often convenient to have simpler
criteria for $\diffeo$ being a groupoid homomorphism.

\blem
\label{lem:graphokrit}
Let $\diffeo : M \nach M$ be some map.
Consider the following statements:
\bnum4
\item
\label{pkt:diffedgepath}
$\diffeo$ maps differentiable edges to paths.
\item
\label{pkt:edgeedge}
$\diffeo$ maps edges to edges.
\item
\label{pkt:hyphhyph}
$\diffeo$ maps hyphs to hyphs.
\item
\label{pkt:pathpath}
$\diffeo$ maps paths to paths.
\item
\label{pkt:groupo}
$\diffeo$ induces a groupoid homomorphism on $\Pf$.
\item
\label{pkt:inj}
$\diffeo$ is injective.
\enum
Then we have the following implications:

\bgl
& \ref{pkt:diffedgepath} \breitrel\aequ 
  \ref{pkt:pathpath} \breitrel\aequ \ref{pkt:groupo} & \\
& \ref{pkt:edgeedge} \breitrel\impliz \ref{pkt:pathpath} & \\
& \ref{pkt:hyphhyph} \breitrel\impliz \ref{pkt:pathpath} & \\
\egl

If $M$ is connected, then even
\bgl
& \ref{pkt:edgeedge}  \breitrel\impliz  \ref{pkt:inj} & \\
\egl

If $\diffeo$ is injective (\ref{pkt:inj}), then 

\bgl
& \ref{pkt:diffedgepath} \breitrel\aequ   
  \ref{pkt:edgeedge} \breitrel\aequ 
  \ref{pkt:hyphhyph} \breitrel\aequ
  \ref{pkt:pathpath} \breitrel\aequ \ref{pkt:groupo}  
\egl

\elem
Note that, by definition,
\ref{pkt:groupo} $\impliz$ \ref{pkt:pathpath} $\impliz$
\ref{pkt:diffedgepath}
\bpf
\bvl
\iitem[\ref{pkt:diffedgepath} $\impliz$ \ref{pkt:pathpath}]
Trivial, since (up to the parametrization)
each path is a product of piecewise differentiable edges and trivial paths.
\iitem[\ref{pkt:edgeedge} $\impliz$ \ref{pkt:pathpath}]
Trivial as well.
\iitem[\ref{pkt:hyphhyph} $\impliz$ \ref{pkt:pathpath}]
Given a path $\gamma$, choose \cite{paper3}
a hyph $\hyph$, such that $\gamma$ is a path
in $\hyph$. By assumption, $\diffeo\circ\hyph$ is a hyph again.
$\diffeo\circ\gamma$ is now a product of elements in $\diffeo\circ\hyph$,
their inverses and trivial paths, hence it is a path as well.
\iitem[\ref{pkt:pathpath} $\impliz$ \ref{pkt:groupo}]
First, we have to check whether $\diffeo$ induces a well-defined
mapping from $\Pf$ to $\Pf$. Let $\gamma, \delta \in \Pfgen$. 
Hence $\diffeo\circ\gamma$ and $\diffeo\circ\delta$ are in $\Pfgen$
again.
If $\gamma$ and $\delta$ coincide up to parametrization, then
also $\diffeo\circ\gamma$ and $\diffeo\circ\delta$ do so. 
If there are $\gamma_1,\gamma_2,\varepsilon \in \Pfgen$
with $\gamma = \gamma_1\gamma_2$ and 
$\delta = \gamma_1\varepsilon \varepsilon^{-1}\gamma_2$,
then similarly
\bgl
\diffeo\circ\gamma & = & (\diffeo\circ\gamma_1) \: (\diffeo\circ\gamma_2)
\egl
and
\bgl
\diffeo\circ\delta & = & (\diffeo\circ\gamma_1) \: (\diffeo\circ\varepsilon) \: 
                      (\diffeo\circ\varepsilon^{-1}) \: (\diffeo\circ\gamma_2) \\
                   & = & (\diffeo\circ\gamma_1) \: (\diffeo\circ\varepsilon) \: 
                      (\diffeo\circ\varepsilon)^{-1} \: (\diffeo\circ\gamma_2);
\egl
hence, $[\diffeo\circ\gamma] = [\diffeo\circ\delta]$. This shows
that equivalent paths are mapped to equivalent paths.
Next, if two paths are composable, their images are composable as well.
Now, the homomorphy property is clear.
\iitem[\ref{pkt:edgeedge} $\impliz$ \ref{pkt:inj}]
Let $x$ and $y$ be two distinct points
in $M$ with $\diffeo(x) = \diffeo(y)$. By connectedness, 
there is an edge $\gamma$
in $M$ running through $x$ and $y$, where at least one of these points
is not an endpoint of $\gamma$. Now the image of $\gamma$ w.r.t.\ $\diffeo$
is not an edge.
\iitem[\ref{pkt:diffedgepath} $\impliz$ \ref{pkt:edgeedge}]
Follows since each path, hence each edge is piecewise differentiable
and since $\diffeo$ is one-to-one.
\iitem[\ref{pkt:edgeedge} $\impliz$ \ref{pkt:hyphhyph}]
By assumption, every hyph is mapped to a finite sequence of edges. 
Observe again that, by injectivity, 
$\diffeo$-images of edges (or segments of them) 
can coincide up to the parametrization only if the origins do so.
Hence, free points are mapped to free points, making images
of hyphs hyphs again.
\qed
\evl
\epf
We remark that, in general, \ref{pkt:pathpath} neither implies
\ref{pkt:hyphhyph} nor \ref{pkt:edgeedge}

\bcorr
A bijection $\diffeo$ is a graphomorphism 
iff both $\diffeo$ and $\diffeo^{-1}$
map edges to paths.
\ecorr
If only $\diffeo$ maps edges to paths, then $\diffeo$ may fail
to be a graphomorphism. In fact, assume that we are working in the 
$C^1$ class on $M = \R^n$ and do not consider embedded paths only. 
Let $\diffeo$ be the homeomorphism mapping $x\in\R^n$ to
$\norm{x} \: x$. Of course, $\diffeo$ maps edges to paths.
Its inverse $\diffeo^{-1} (x) = \norm{x}^{-\einhalb} \: x$, however,
does not. For example, the straight line $\gamma(t) = t \:e$ with $e \in \R^n$
having norm $1$ is mapped to $(\diffeo^{-1} \circ \gamma)(t) = \sqrt t \: e$
being not differentiable at $t = 0$.

Finally, we have
\blem
\label{lem:graphos_preserve}
Every graphomorphism on $M$ maps (complete) hereditary subsets of
$\Pfgen$ into (complete) hereditary ones.
\elem


\section{Main Results}
\label{sect:main}
In this section we are going to provide the general scheme, comprising
several transformations on $\Ab$. Let us very briefly recall the two most
important ones -- the gauge transforms and the diffeomorphisms. 

The set $\Gb$ of gauge transforms consists 
of all maps $\qg$ from $M$ to $\LG$ acting on $\Ab$
by%
\footnote{Until Theorem \ref{thm:cont+masz(qfa)}, we simply drop 
the square brackets in expressions like $h_\qa([\gamma])$,
since confusions are not to be expected. That all that is well defined 
will be proven below.}
\bgl
h_{\qa\circ\qg} (\gamma) 
 & := & \qg(\gamma(0))^{-1} \: h_\qa(\gamma) \: \qg(\gamma(1))
\breitrel{\breitrel{\text{for all $\gamma\in\Pfgen$}}}
\egl\noindent
and is given the product topology on $\Maps(M,\LG) \iso \LG^M$.
The action of diffeomorphisms 
on $M$ can be lifted to an action on $\Pfgen$ (and $\Pf$),
which again can be lifted to an action on $\Ab$. 
In fact, each diffeomorphism $\diffeo$ defines
a map from $\Ab$ to $\Maps(\Pf,\LG)$,
again denoted by $\diffeo$,
via
\bgl\
h_{\diffeo (\qa)}(\gamma) & := & h_\qa(\diffeo^{-1}\circ\gamma) 
\breitrel{\breitrel{\text{for all $\gamma\in\Pfgen$.}}}
\egl\noindent

As we will see also for other examples like the Weyl transformations later,
there are two typical features characterizing these
transformations $\qfa$ on $\Ab$. First, 
there is given some set $\adm$ of elementary paths
(we had $\adm = \Pfgen$ in the examples above, but have to reduce this
set, e.g., for Weyl transformations).
And, second, the modified parallel transport along some 
path in $\adm$ depends always only on 
\bunum
\item
the original parallel transport for some, possibly 
different path in $\adm$, and
\item
some conjugation-like multiplication
by some group elements depending on the path only.
\eunum
In other words, we have
\bgl
h_{\qfa(\qa)}(\gamma) 
  & = & f_1(\gamma)^{-1} \: h_\qa(\diffeo (\gamma)) \:
        f_2(\gamma)
\egl\noindent
with some functions $f_1, f_2 : \adm \nach \LG$ and some mapping 
$\diffeo : \adm \nach \adm$.
Let us now check criteria to make such a general transformation
well defined on $\Ab$. First of all, $\diffeo$ should 
map edges to edges and extend to a well-defined
mapping from $[\adm]$ to $[\adm]$. 
For simplicity, let us assume additionally that $\diffeo$ is even induced by a 
graphical homomorphism 
and let (just for the next few lines) $\LG$ have trivial center.
We get
\bgl
        f_1(\gamma^{-1})^{-1} \: h_\qa(\diffeo \circ \gamma)^{-1} \:
        f_2(\gamma^{-1})
  & = & 
        f_1(\gamma^{-1})^{-1} \: h_\qa(\diffeo \circ \gamma^{-1}) \:
        f_2(\gamma^{-1}) 	\\
  & = & h_{\qfa(\qa)}(\gamma^{-1}) \\
  & = & h_{\qfa(\qa)}(\gamma)^{-1} \\
  & = & f_2(\gamma)^{-1} \: h_\qa(\diffeo \circ \gamma)^{-1} \:
        f_1(\gamma) \\
\egl\noindent
for all $\gamma\in\adm$.  This implies%
\footnote{Let $G$ be a group and $H$ a subgroup of $G$. Let, moreover, 
$a_1, a_2, b_1, b_2$ be in $G$, 
such that $a_1 h a_2 = b_1 h b_2$ for all $h\in H$. 
Then $a_1 \in b_1 Z_G(H)$ and $a_2 \in Z_G(H) b_2$.
In fact, we have $h^{-1} b_1^{-1} a_1 h = b_2 a_2^{-1} = b_1^{-1} a_1$
for all $h\in H$, where the second equality follows for $h = e_G$.
Hence, $b_1^{-1} a_1$ and $H$ commute.}
$f_1(\gamma) = f_2(\gamma^{-1}) =: f(\gamma)$,
since $\pi_\gamma : \Ab \nach \LG$ is surjective and $\LG$ has
trivial center.
Next, if a nontrivial path in $\adm$ can be decomposed into two 
nontrivial paths
$\gamma_1$ and $\gamma_2$ in $\adm$, we get
\bgl
 &   & 
       f(\gamma_1 \gamma_2)^{-1} \: h_\qa(\diffeo \circ \gamma_1) \:
       h_\qa(\diffeo \circ \gamma_2) \: f((\gamma_1 \gamma_2)^{-1}) \\
 & = & f(\gamma_1 \gamma_2)^{-1} \: h_\qa(\diffeo \circ (\gamma_1\gamma_2)) \:
       f((\gamma_1 \gamma_2)^{-1}) \\
 & = & h_{\qfa(\qa)}(\gamma_1 \gamma_2) \\
 & = & h_{\qfa(\qa)}(\gamma_1) \: h_{\qfa(\qa)}(\gamma_2) \\
 & = & f(\gamma_1)^{-1} \: h_\qa(\diffeo \circ \gamma_1) \: f(\gamma_1^{-1}) \:\:
       f(\gamma_2)^{-1} \: h_\qa(\diffeo \circ \gamma_2) \: f(\gamma_2^{-1}).
\egl\noindent
Typically, $\gamma_1$ and $\gamma_2$ form a hyph. Hence,
the corresponding parallel transports can be assigned independently.
By the triviality of the center of $\LG$, we get    
$f(\gamma_1 \gamma_2) = f(\gamma_1)$ and $f(\gamma_1^{-1}) = f(\gamma_2)$.

\extrazeile[0.8]
This motivates (now back to the case of an arbitrary connected Lie group $\LG$)
\bdf
Let $\adm$ be some hereditary subset of $\Pfgen$.

Then a map $\rrr : \adm \nach \LG$ is called \df{admissible} iff
\bunum
\item
$\rrr(\delta_1) = \rrr(\delta_2)$
for all $\delta_1,\delta_2\in\adm$
with $\delta_1 \BB \delta_2$,
and
\item
$\rrr(\gamma_1^{-1}) = \rrr(\gamma_2)$
for all $\gamma\in\adm$ and all decompositions $\gamma_1\gamma_2$ of $\gamma$.
\eunum
\edf
Now, we may state the main
\bthm
\label{thm:cont+masz(qfa)}
Let $\adm$ be some complete subset of $\Pfgen$,
and let $\diffeo$ be some graphical homomorphism of $M$.
Moreover, let $\rrr : \adm \nach \LG$ 
be some admissible map.

Then we have:
\bnum3
\item
There is a unique continuous map $\qfa : \Ab \nach \Ab$,
such that, for all $\gamma \in \adm$,
\bgl
h_{\qfa(\qa)}([\gamma])  & = & 
\rrr(\gamma)^{-1} \: h_\qa([\diffeo \circ \gamma]) \: \rrr(\gamma^{-1}).
\egl
\item
If $\diffeo$ is injective, then $\qfa$ even
preserves
the Ashtekar-Lewandowski measure $\mu_0$. Hence, 
the induced operator
on $\bound(L_2(\Ab,\mu_0))$ is well defined and unitary.
Moreover, the pull-back  
$\qfa^\ast : C(\Ab) \nach C(\Ab)$ is an isometry.
\item
If $\diffeo$ is a graphomorphism, then
$\qfa$ is even a homeomorphism. 
\enum
\ethm
Recall that, for compact $\LG$,
the Ashtekar-Lewandowski measure $\mu_0$ is the unique
regular Borel measure on $\Ab$ whose push-forward $(\pi_\hyph)_\ast\mu_0$ 
to $\Ab_\hyph \iso \LG^{\elanz\hyph}$ 
coincides with the Haar measure there for every hyph $\hyph$. 
\bpf
\bnum3
\item
$\qfa$ exists uniquely and is continuous.
\bunum
\item
Define $\qwert : \Ab \nach \Maps(\adm,\LG)$ by%
\footnote{From now on, we will drop the square brackets 
in all $h_\qa([\ldots])$.}
\bgl
\bigl(\qwert(\qa)\bigr)(\gamma)  & = & 
\rrr(\gamma)^{-1} \: h_\qa (\diffeo \circ \gamma) \: \rrr(\gamma^{-1}).
\egl
\item
First we show that $\qwert(\qa)$ is indeed in $\Germ(\adm,\LG)$
for all $\qa\in\Ab$. 

In fact,
for all $\gamma\in\adm$ and all decompositions $\gamma_1\gamma_2$ of $\gamma$,
we have
\bgl
\bigl(\qwert(\qa)\bigr)(\gamma^{-1})
 & = & \rrr(\gamma^{-1})^{-1} \: h_\qa (\diffeo \circ \gamma^{-1}) \: \rrr(\gamma) \\
 & = & \bigl(\rrr(\gamma)^{-1} \: h_\qa (\diffeo \circ \gamma) \: \rrr(\gamma^{-1})\bigr)^{-1} 
 \breitrel= \bigl(\qwert(\qa)(\gamma)\bigr)^{-1}
\egl
and
\bgl
\bigl(\qwert(\qa)\bigr)(\gamma)
 & = & \rrr(\gamma)^{-1} \: h_\qa (\diffeo \circ \gamma) \: \rrr(\gamma^{-1}) \\
 & = & \rrr(\gamma_1\gamma_2)^{-1} \: h_\qa (\diffeo \circ \gamma_1) \:
       h_\qa (\diffeo \circ \gamma_2) \: \rrr(\gamma_2^{-1}\gamma_1^{-1}) \\
 & = & \rrr(\gamma_1)^{-1} \: h_\qa (\diffeo \circ \gamma_1) \:
       \rrr(\gamma_1^{-1}) \: \rrr(\gamma_2)^{-1} \:
       h_\qa (\diffeo \circ \gamma_2) \: \rrr(\gamma_2^{-1}) \\
 & = & \bigl(\qwert(\qa)\bigr)(\gamma_1) \: \bigl(\qwert(\qa)\bigr)(\gamma_2).
\egl
Here, we used that $\gamma_1 \BB \gamma_1 \gamma_2$
and $\gamma_2^{-1} \gamma_1^{-1} \BB \gamma_2^{-1}$.
\item
Next, observe that for every fixed $\gamma\in\adm$,
\bgl
\bigl(\qwert(\qa)\bigr)(\gamma)
 & = & \rrr(\gamma)^{-1} \: h_{\qa}(\diffeo \circ \gamma) \: \rrr(\gamma^{-1})
 \breitrel\ident \rrr(\gamma)^{-1} \: \pi_{\diffeo \circ \gamma} (\qa) \: \rrr(\gamma^{-1})
\egl
depends continuously on $\qa$, by definition of the projective-limit topology
on $\Ab$.
\item
Now, by Proposition \ref{prop:qfa_cont},
$\qfa := [\neuzh{\qwert(\cdot)}] : \Ab \nach \Ab$ is continuous, whereas
for $\gamma\in\adm$
\bgl
\!\! h_{\qfa(\qa)}(\gamma) 
 \breitrel\ident \bigl(\qfa(\qa)\bigr)([\gamma]) 
 & = & [\neuzh{\qwert(\qa)}] ([\gamma]) 
 \breitrel= \rrr(\gamma)^{-1} \: h_\qa (\diffeo \circ \gamma) \: \rrr(\gamma^{-1}).
\!\!
\egl
The uniqueness of $\qfa$ follows from the completeness of $\adm$
and Lemma \ref{lem:coinc_krit}.
\eunum
\item
$\qfa$ preserves the Ashtekar-Lewandowski measure and
has isometric pull-back for injective $\diffeo$.
\bunum
\item

In fact, let $\hyph$ be an arbitrary, but fixed hyph. 
By completeness, there is some hyph $\hyph'\geq\hyph$ with 
$Y'$ edges, such that
every $\gamma_i\in\hyph'$ is in $\adm$:
Indeed, first decompose each path $\gamma\in\hyph$ into a product of 
paths in $\adm$. Collect the paths used there, in some set $\gc \geq \hyph$.
Since possibly $\gc$ is not a hyph again, decompose, if necessary, 
the paths in $\gc$ further to get a hyph 
$\hyph'\geq\gc\geq\hyph$ \cite{paper3}. By construction,
$\gc$ is contained in $\adm$. Now, by heredity of $\adm$, so does
$\hyph'$.

By construction, we have 
\bgl
\pi_{\hyph'} \circ \qfa  & = &  
   (\qfa_{\gamma_1} \kreuz \cdots \kreuz \qfa_{\gamma_{\Hyph'}}) \circ \pi_{\diffeo\circ\hyph'}
\egl
with 
$\qfa_\gamma (g) := 
  \rrr(\gamma)^{-1} \: g \: \rrr(\gamma^{-1})$ for $\gamma\in\adm$.
In other words, each $\qfa_\gamma$ consists of a left and a right translation,
whence the Haar measure on $\LG$ is $\qfa_\gamma$-invariant.
\item
Since $\pi_\hyph^{\hyph'} \circ \pi_{\hyph'} = \pi_\hyph$ with continuous
$\pi_\hyph^{\hyph'} : \Ab_{\hyph'} \nach \Ab_\hyph$,
since $(\pi_{\hyph'})_\ast \mu_0$ is the $Y'$-fold product of the 
Haar measure on $\LG$ and since $\diffeo \circ \hyph'$ is
again a hyph of $Y'$ edges by Lemma \ref{lem:graphokrit}, we get
\bgl
(\pi_\hyph)_\ast (\qfa_\ast \mu_0) 
 & = & (\pi_\hyph^{\hyph'})_\ast (\pi_{\hyph'} \circ \qfa)_\ast \mu_0  \\
 & = & (\pi_\hyph^{\hyph'})_\ast (\qfa_{\gamma_1} \kreuz \cdots \kreuz \qfa_{\gamma_{\Hyph'}})_\ast
       (\pi_{\diffeo\circ\hyph'})_\ast \mu_0 \\
 & = & (\pi_\hyph^{\hyph'})_\ast (\qfa_{\gamma_1} \kreuz \cdots \kreuz \qfa_{\gamma_{\Hyph'}})_\ast
       \mu_\Haar^{\Hyph'} \\
 & = & (\pi_\hyph^{\hyph'})_\ast \mu_\Haar^{\Hyph'} \\
 & = & (\pi_\hyph^{\hyph'})_\ast (\pi_{\hyph'})_\ast \mu_0 \\
 & = & (\pi_\hyph)_\ast \mu_0.
\egl
Since finite regular Borel measures on $\Ab$ 
coincide iff their push-forwards w.r.t.\ 
all $\pi_\hyph$ coincide, we get the assertion.
\item
Let $f$ be some cylindrical function on $\Ab$ w.r.t.\ $\hyph$, i.e., we
have $f = f_\hyph \circ \pi_\hyph$ for some continuous $f_\hyph$ on $\LG^Y$.
Now, 
\bgl 
\qfa^\ast f & \ident & f \circ \qfa \breitrel=
  f_\hyph \circ \pi_\hyph^{\hyph'} \circ
      (\qfa_{\gamma_1} \kreuz \cdots \kreuz \qfa_{\gamma_{\Hyph'}}) 
      \circ \pi_{\diffeo\circ\hyph'}.
\egl
Since $\pi_{\widetilde\hyph}$ is surjective for all hyphs $\widetilde\hyph$,
since $\qfa_{\gamma_1} \kreuz \cdots \kreuz \qfa_{\gamma_{\Hyph'}}$ 
is surjective and since $\pi_\hyph^{\hyph'}$
is surjective, we have
$\supnorm{\qfa^\ast f} = \supnorm{f_\hyph} = \supnorm{f}$.
Since cylindrical functions are dense in $C(\Ab)$ and since $\qfa^\ast$ is
continuous, we get the assertion.
\eunum
\item
$\qfa$ is a homeomorphism if $\diffeo$ is a graphomorphism.
\bunum
\item
Observe that $\diffeo(\adm)$ is complete by 
Lemma \ref{lem:graphos_preserve}.
Now, define $\rrr' : \diffeo(\adm) \nach \LG$
by $\rrr'(\gamma) := \rrr(\diffeo^{-1} \circ \gamma)^{-1}$.
It is easy to check that $\rrr'$ is admissible w.r.t.\ $\diffeo(\adm)$.
As already proven above, there is a unique continuous map
$\qfa' : \Ab \nach \Ab$ with 
\bgl
h_{\qfa'(\qa)}(\gamma)  & = & 
\rrr'(\gamma)^{-1} \: h_\qa(\diffeo^{-1} \circ \gamma) \: \rrr'(\gamma^{-1})
\egl
for all $\gamma\in\diffeo(\adm)$. Altogether, this gives
\bgl
 &   & h_{\qfa'(\qfa(\qa))}(\gamma)  \\
 & = & \rrr'(\gamma)^{-1} \: 
          h_{\qfa(\qa)} (\diffeo^{-1} \circ \gamma) \: 
          \rrr'(\gamma^{-1}) \\
 & = & \rrr'(\gamma)^{-1} \: 
          \rrr(\diffeo^{-1}\circ \gamma)^{-1} \: 
          h_\qa(\diffeo \circ \diffeo^{-1} \circ \gamma) \:
          \rrr((\diffeo^{-1} \circ \gamma)^{-1}) \:
          \rrr'(\diffeo\circ\gamma^{-1}) \\
 & = & h_\qa(\gamma) 
\egl
for all $\gamma\in\diffeo(\adm)$.
The completeness of $\diffeo(\adm)$ and Lemma \ref{lem:coinc_krit} prove
$\qfa' \circ \qfa = \ido_\Ab$. Analogously, one shows 
$\qfa \circ \qfa' = \ido_\Ab$.
\qed
\eunum
\enum\epf

We get immediately
\bcorr
\label{corr:cont(qfa)voll}
Let $\adm$ be some complete subset of $\Pfgen$,
let $\diffeo$ be some graphical homomorphism of $M$,
let $Y$ be some topological space
and let $\rrr : \adm \kreuz Y \nach \LG$
be some map, such that 
\bunum
\item
$\rrr(\cdot,y) : \adm \nach \LG$ is admissible for all $y\in Y$, and
\item
$\rrr(\gamma,\cdot) : Y \nach \LG$ is continuous for all $\gamma\in\adm$.
\eunum
Then there is a unique map $\qfa : \Ab \kreuz Y \nach \Ab$ with 
\bgl
h_{\qfa(\qa,y)}(\gamma) 
  & = & \rrr(\gamma,y)^{-1} \: h_{\qa}(\diffeo\circ\gamma) \: 
                 \rrr(\gamma^{-1},y) 
\egl
for all $\gamma \in \adm$. Moreover, $\qfa$ is continuous.
\ecorr

\section{Applications}
\label{sect:examples}

Let us consider three basic examples.
\bcorr[Gauge Transforms]
There is a unique map $\Theta : \Ab \kreuz \Gb \nach \Ab$,
such that
\bgl
h_{\Theta(\qa,\qg)} (\gamma) 
 & = & \qg(\gamma(0))^{-1} \: h_\qa(\gamma) \: \qg(\gamma(1))
\breitrel{\breitrel{\text{for all $\gamma\in\Pfgen$.}}}
\egl\noindent
$\Theta$ is continuous. Moreover, 
$\Theta_\qg := \Theta(\cdot,\qg) : \Ab \nach \Ab$ 
is a homeomorphism and preserves
the Ashtekar-Lewandowski measure
for each $\qg\in\Gb$. The inverse of $\Theta_\qg$ is given
by $\Theta_{\qg^{-1}}$.
\ecorr
Usually, one writes $\qa \circ \qg$ instead of $\Theta(\qa,\qg)$.
\bpf
Set $\adm := \Pfgen$ and $\diffeo := \ido_M$. Moreover,
set $Y := \Gb$, and define $\rrr(\gamma,\qg) := \qg(\gamma(0))$.
Of course, $\rrr$ fulfills 
the requirements of Corollary \ref{corr:cont(qfa)voll}.
The assertion now follows from
$\qg(\gamma^{-1}(0)) = \qg(\gamma(1))$,
Theorem \ref{thm:cont+masz(qfa)} and Corollary \ref{corr:cont(qfa)voll}.
\qed
\epf

\bcorr[Diffeomorphisms]
There is a unique map $\Theta : \Ab \kreuz \Grapho(M) \nach \Ab$, such that
\bgl\
h_{\Theta(\qa,\diffeo)}(\gamma) & = & h_\qa(\diffeo^{-1}\circ\gamma) 
\breitrel{\breitrel{\text{for all $\gamma\in\Pfgen$.}}}
\egl\noindent
Moreover, 
$\Theta_\diffeo := \Theta(\cdot,\diffeo) : \Ab \nach \Ab$ 
is a homeomorphism and preserves
the Ashtekar-Lewandowski measure
for each graphomorphism $\diffeo\in\Grapho(M)$. 
The inverse of $\Theta_\diffeo$ is given
by $\Theta_{\diffeo^{-1}}$.
\ecorr
Usually, one writes $\diffeo(\qa)$ instead of $\Theta(\qa,\diffeo)$.
\bpf
Define $\adm := \Pfgen$ and $\rrr(\gamma) := e_\LG$ for all $\gamma\in\LG$. 
Theorem \ref{thm:cont+masz(qfa)} gives the proof with inverted $\diffeo$.
\qed
\epf

\bcorr[Weyl transformations]
Let $S$ be a quasi-surface, and let $\adm$ consist of all edges
and trivial paths 
$\gamma$ being $S$-external (i.e., $\inter\gamma \cap S = \leeremenge$)
or $S$-internal (i.e., $\inter\gamma \teilmenge S$). Moreover,
let $\sigma_S$ be some intersection function for $S$.

Then there is a unique map 
$\qfa^{S,\sigma_S} : \Ab \kreuz \Maps(M,\LG) \nach \Ab$,
such that

\bgl
h_{\qfa^{S,\sigma_S}(\qa,d)} (\gamma)  & \!\!\! = \!\!\! & 
 \begin{cases}
    d(\gamma(0))^{\sigma^\ausl_S(\gamma)} \: 
    h_\qa(\gamma) \: 
    d(\gamma(1))^{\sigma^\einl_S(\gamma)}
    & \text{if $\gamma$ is $S$-external } \\
    \phantom{d(\gamma(0))^{\sigma^\ausl_S(\gamma)}} \:
    h_\qa(\gamma) \: 
    & \text{if $\gamma$ is $S$-internal } 
  \end{cases}.
\egl
Moreover, 
the map 
$\qfa^{S,\sigma_S}_d  :  \Ab \nach \Ab$,
given
by $\qfa^{S,\sigma_S}_d(\qa) := \qfa^{S,\sigma_S} (\qa,d)$, 
is a homeomorphism and preserves
the Ashtekar-Lewandowski measure for each $d\in\Maps(M,\LG)$.
The inverse of $\qfa^{S,\sigma_S}_d$ is given 
by $\qfa^{S,\sigma_S}_{d^{-1}}$.
Finally, if $\Maps(M,\LG) \iso \LG^M$ is given the product topology, then
$\qfa^{S,\sigma_S}$ is continuous.

\ecorr
For the definition of quasi-surfaces and intersection functions, see 
\cite{paper18}. Moreover, note that the Weyl operators are
the pull-backs of the corresponding 
Weyl transformations $\qfa^{S,\sigma_S}_d$ to
$C(\Ab)$ (or their induced action on $L_2(\Ab,\mu_0)$).
\bpf
\bunum
\item
$\adm$ is complete.

This follows since $S$ is a quasi-surface, i.e.,
every edge (hence any finite path) can be 
decomposed (up to the parametrization) into a product 
of edges and trivial paths being $S$-external or $S$-internal.\ \cite{paper18}
\item
$\qfa^{S,\sigma_S}$ exists uniquely and 
is continuous for the product topology on $\Maps(M,\LG)$.

Let $\diffeo$ be the identity on $M$, and let $Y := \Maps(M,\LG)$.
Moreover, let
\bgl
\rrr(\gamma,d) & := & 
 \begin{cases}
    d(\gamma(0))^{-\sigma^\ausl_S(\gamma)}    
    & \text{if $\gamma$ is $S$-external } \\
    e_\LG \: 
    & \text{if $\gamma$ is $S$-internal } 
  \end{cases}.
\egl
The only nontrivial property of $\rrr$ 
in Corollary \ref{corr:cont(qfa)voll} to be checked is 
$\rrr(\gamma_1^{-1},d) = \rrr(\gamma_2,d)$ for 
decompositions $\gamma_1 \gamma_2$ of $S$-external $\gamma$.
Observe, however, that here 
$\gamma_1^{-1} (0) \ident \gamma_1(1) \ident \gamma_2(0)$ is not contained
in $S$, hence 
$\rrr(\gamma_1^{-1},d) = e_\LG = \rrr(\gamma_2,d)$.
The claim now follows from 
$\sigma^\ausl_S(\gamma) + \sigma^\einl_S(\gamma^{-1}) = 0$
and Corollary \ref{corr:cont(qfa)voll}.
\item
$\qfa^{S,\sigma_S}_d$ is a homeomorphism and 
leaves $\mu_0$ invariant.

This now follows from Theorem \ref{thm:cont+masz(qfa)}.
\qed
\eunum
\epf

Finally, we may use the theorem above to 
rederive and extend results known from \cite{paper3}. 

\bcorr
Let $N$ be some set of points in $M$ having no accumulation point,
and let $\adm$ be the set of all paths that are $N$-external or trivial.
For every $x\in N$, let $E_x$ be some set of edges starting at $x$,
such that 
\bnum2
\item
$N \cap \im E_x = \{x\}$ and 
\item
$\gamma_1 \BB \gamma_2$ for $\gamma_1,\gamma_2 \in E_x$ 
implies $\gamma_1 = \gamma_2$.
\enum
Define $E$ to be the union of all $E_x$.
Moreover, let $f : E \nach \LG$ be some function
and define for $\gamma\in\adm$
\bgl
\rrr(\gamma,\qa) & := & 
 \begin{cases} 
    h_\qa(e) \: f(e)^{-1}
          & \text{ if $\gamma \BB e$ for some $e \in E$} \\
    e_\LG & \text{ otherwise}
 \end{cases}.
\egl

Then there is a unique map $\qfa : \Ab \nach \Ab$, 
such that 
\bgl
h_{\qfa(\qa)}(\gamma) 
  & = & \rrr(\gamma,\qa)^{-1} \: h_{\qa}(\gamma) \: \rrr(\gamma^{-1},\qa) 
\egl
for all $\gamma\in\adm$.
Moreover, $\qfa$ is continuous.
\ecorr
\bpf
\bunum
\item
$\adm$ is complete.

Since $N$ does not contain accumulation points, every path can
be decomposed into finitely many 
subpaths either not containing any point of $N$ in
their interior or being trivial. The heredity is clear.
\item
$\rrr$ is well defined.

If there exist $e_1, e_2 \in E$ with $\gamma \BB e_1$ and $\gamma \BB e_2$,
then $e_1 \BB e_2$ and $e_1,e_2 \in E_{\gamma(0)}$, hence $e_1 = e_2$
by assumption.
\item
$\rrr(\cdot,\qa)$ is admissible for all $\qa\in\Ab$.

Let $\delta_1,\delta_2\in\adm$ with $\delta_1 \BB \delta_2$.
If they are trivial, then even $\delta_1 = \delta_2$. Otherwise, they are 
$N$-external. Then one of them starts as some $e\in E$ iff the other 
does. Hence, in both cases, $\rrr(\delta_1,\qa) = \rrr(\delta_2,\qa)$ 
for all $\qa\in\Ab$.
Let now $\gamma$ be in $\adm$ and $\gamma_1\gamma_2$ be a decomposition of 
$\gamma$. If $\gamma$ is $N$-external, then,
in particular, neither $\gamma_1^{-1}$ nor $\gamma_2$ starts as any $e\in E$, 
since $\gamma_1^{-1}(0) \ident \gamma_2(0) = \gamma(\einhalb) \nichtin N$. 
Hence, we have $\rrr(\gamma_1^{-1},\qa) = e_\LG = \rrr(\gamma_2,\qa)$. The case
of trivial $\gamma$ is clear.
\item
$\qfa$ is continuous.

By Corollary \ref{corr:cont(qfa)voll}, it is sufficient to show
that $\rrr(\gamma,\qa)$ is continuous in $\qa$.
This however follows, because, by construction, 
there is at most one $e \in E$ with $\gamma \BB e$.
\qed
\eunum
\epf

\neueseite
\noindent
Note, that, in general, $\qfa$ does not leave the Ashtekar-Lewandowski measure
invariant. In fact, we have $h_{\qfa(\qa)}(e) = f(e)$
for every $e\in E$. Therefore, 
$\qfa(\Ab) \teilmenge \pi_e^{-1}(\{f(e)\})$,
hence 
\bgl
 0 \breitrel\leq \mu_0(\qfa(\Ab)) 
 \breitrel\leq \mu_0 \bigl(\pi_e^{-1}(\{f(e)\})\bigr) 
 \breitrel= \mu_\Haar(\{f(e)\}) \breitrel= 0,
\egl\noindent
unless $\LG$ is trivial.

\section{Acknowledgements}
The author thanks Garth Warner for fruitful discussions and,
in particular, for notifying a mistake in 
a proof in \cite{paper18}%
\footnote{Version {\sf math-ph/0407006 v1}, dated July 4, 2004}. 
Curing this problem has led to the
present paper.
The author has been supported in part by NSF grant PHY-0090091.



\begin{thebibliography}{1}

\bibitem{a28}
{Abhay Ashtekar and Jerzy Lewandowski: Differential geometry on the space of
  connections via graphs and projective limits. {\it J. Geom. Phys.} {\bf 17}
  (1995) {191--230}. {\sf e-print:\ hep-th/9412073}.}

\bibitem{paper3}
{Christian Fleischhack: Hyphs and the Ashtekar-Lewandowski Measure. {\it J.
  Geom. Phys.} {\bf 45} (2003) {231--251}. {\sf e-print:\ math-ph/0001007}.}

\bibitem{paper18}
{Christian Fleischhack: Representations of the Weyl Algebra in Quantum
  Geometry. {\sf e-print:\ math-ph/0407006}.}

\bibitem{paper2+4}
{Christian Fleischhack: Stratification of the Generalized Gauge Orbit Space.
  {\it Commun. Math. Phys.} {\bf 214} (2000) {607--649}. {\sf e-print:\
  math-ph/0001006, math-ph/0001008}.}

\end{thebibliography}
\end{document}